\documentclass[showpacs,10pt,twocolumn,prb]{revtex4-1}

\usepackage{amsmath}
\usepackage{amssymb}
\usepackage{graphicx}
\usepackage{amssymb}
\usepackage{graphics}
\usepackage{epsfig}
\usepackage{CJK}
\usepackage{color}

\begin{document}

\title{Interlayer electronic transport in CaMnBi$_{2}$ antiferromagnet}
\author{Aifeng Wang,$^{1}$, D. Graf,$^{2}$ Lijun Wu,$^{2}$ Kefeng Wang,$^{3,\ast}$ E. Bozin,$^{1}$ Yimei Zhu$^{1}$ and C. Petrovic$^{1}$}
\affiliation{$^{1}$Condensed Matter Physics and Materials Science Department, Brookhaven National Laboratory, Upton, New York 11973, USA\\
$^{2}$National High Magnetic Field Laboratory, Florida State University, Tallahassee, Florida 32306-4005, USA}

\date{\today}

\begin{abstract}
We report interlayer electronic transport in CaMnBi$_{2}$ single crystals. Quantum oscillations and angular magnetoresistance suggest coherent electronic conduction and valley polarized conduction of Dirac states. Small cyclotron mass, large mobility of carriers and nontrivial Berry's phase are consistent with the presence of Dirac fermions on the side wall of the warped cylindrical Fermi surface. Similar to SrMnBi$_{2}$ that features an anisotropic Dirac cone, our results suggest that magnetic field-induced changes in the interlayer conduction are also present in layered bismuth-based materials with zero-energy line in momentum space created by the staggered alkaline earth atoms.
\end{abstract}
\maketitle

\section{INTRODUCTION}

Similar to graphene and topological insulators, ternary AMnBi$_2$ crystals (A = alkaline earth, such as Ca,Sr or Ba) also host quasi-2D Dirac states \cite{ParkSr,KefengSr,KefengCa,LijunBa}. The Dirac cone in SrMnBi$_2$ is strongly anisotropic due to spin orbit coupling, which is essential ingredient for the magnetic valley control and valley-polarized interlayer current \cite{LeeG,JoYJ,ZhuZW,KuchlerR}. There might be multiple degenerate 'valleys' (conduction band minima) for the carriers to occupy in the electronic structure of certain crystal lattice. Such degeneracy can be lifted in a controllable way; i.e. magnetic valley control is a degeneracy controlled by magnetic field. The valley degree of freedom can be used to develop electronic devices \cite{ZhuZW, Xiao D}. First-principle calculations and angle-resolved photoemission (ARPES) measurements indicate that the anisotropy of the Dirac cone is determined by the local arrangement of Sr/Ca surrounding the Bi square net \cite{LeeG,FengY}. The stacking configuration of the two alkaline earth atomic layers above and below the Bi square net is different for SrMnBi$_2$ and CaMnBi$_2$, creating an anisotropic Dirac cone (SrMnBi$_{2}$) or a zero-energy line in momentum space (CaMnBi$_{2}$) \cite{LeeG}. Therefore, it is of interest to probe the interlayer conduction in  CaMnBi$_2$.

The valley control in SrMnBi$_2$ is realized through field dependent coherent interlayer conduction, sensitive to the curvature of the side wall of the quasi-2D Fermi surface (FS) \cite{JoYJ}. This is similar to for example quasi two-dimensional (2D) organic superconductor magnetoresistance (MR) peak structure when the magnetic field is nearly parallel to the conducting plane. The MR is explained by the open orbits or small closed orbits formed on the side of the warped FS \cite{Hanasaki}. The warped FS is easily detected  by Shubnikov de Haas (SdH) oscillations since the oscillation frequency is determined by the extremal orbit \cite{Shoeneberg}.

In-plane electronic transport reveals that CaMnBi$_2$ is a bad metal with antiferromagnetic transition at $\sim$ 250 K, showing giant magnetoresistance \cite{KefengCa,WangJK,HeJB}. The MR and quantum oscillations results indicate the existence of quasi-2D Dirac fermions in CaMnBi$_2$ \cite{KefengCa}. Negative thermopower suggests dominant electron-type carriers whereas magnetic field suppresses its absolute value, consistent with the presence of Dirac fermions \cite{KefengTTO}.

Here we report the angular dependent MR and SdH in CaMnBi$_2$ when the current runs along the $c$-axis of the crystal. The four fold symmetry of the azimuthal angle dependence of out-of-plane resistivity ($\rho_c$) indicates valley degeneracy contribution of Dirac fermions to interlayer conductivity. Moreover the contribution can be lifted and controlled by the in-plane magnetic field. When compared to SrMnBi$_2$, the out-of-plane resistivity and Hall resistance suggest larger contribution of 3D FS in conductivity. The peak of the MR when the field is in the $ab$ plane and the narrow angle range of the SdH indicate the existence of small closed orbits on the side of the warped FS. The SdH along the $c$ axis features 3 peaks, possibly due to several extremal orbits on the $\alpha$ band and consistent with the angle dependent MR results. Temperature dependence of the SdH in the $ab$ plane reveals small cyclotron mass, large mobility, and nozero Berry's phase on the small closed pocket. That suggests that zero-energy line of Dirac carriers in CaMnBi$_2$ contains small closed orbits \cite{LeeG,FengY}.

\section{EXPERIMENTAL DETAILS}

CaMnBi$_2$ single crystals were grown from high temperature bismuth flux \cite{KefengCa}. Neutron time-of-flight powder diffraction measurement at 300 K was performed on the POWGEN instrument, BL-11A at the Spallation Neutron Source, Oak Ridge National Laboratory. Vanadium can containing 0.5 g of finely pulverized sample was used with in situ sample changer. Powder used in the Neutron experiment is obtained by pulverizing single crystals from same batch of the single crystal used in the interlayer transport experiments. The average structure was assessed through Rietveld refinements to the raw diffraction data using GSAS operated under EXPGUI, utilizing tetragonal space group $P4/nmm$ \cite{Rietveld,Larson,Toby}. Transmission-electron-microscopy (TEM) sample was prepared by crushing the single crystal sample, and then dropping to Lacey carbon grid. High-resolution TEM imaging was performed using the double aberration-corrected JEOL-ARM200CF microscope with a cold-field emission gun operated at 200 kV. Single crystals free of bismuth flux for magnetotransport measurements was obtained by cleaving and cutting the six faces of the cuboid. Magnetotransport measurements up to 9 T were performed using Quantum Design PPMS-9  and up to 35 T at the National High Magnetic Field Laboratory (NHMFL) in Tallahassee. For the out-of-plane resistance measurement, a thick single crystal was cut to produce a needle-like sample with long side along the $c$-axis with about 5$^{\circ}$ uncertainty in order to minimize the contribution of the in-plane resistivity component. This is contrast to the method where voltage contacts are attached on the opposite [001] planes of rectangular crystal \cite{KefengCa,WangXF,EdwardsJ}. Electrical contacts used in resistivity measurements were made to the samples using silver paste to attach Pt wires in a standard four-probe configuration. Temperature dependence of resistivity on three different independently grown crystals from the same batch was reproducible and consistent with each other. Given the sample size, the error introduced by geometry factor can be as high as 18\%.

\section{RESULTS AND DISCUSSIONS}

\begin{figure}[ht]
\centering
\includegraphics[width=0.495\textwidth]{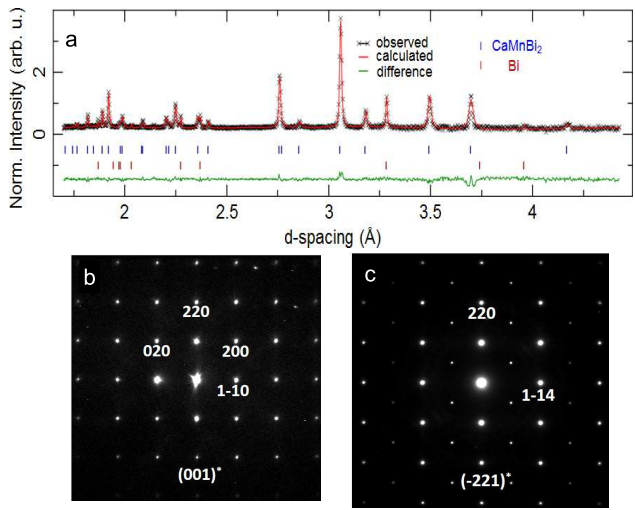}
\caption{(Color online). (a) Structural refinement of neutron powder diffraction data of CaMnBi$_{2}$ at 300 K. Electron diffraction pattern of CaMnBi$_{2}$ viewed along (b) [001] and (c) [-221] direction. Reflection conditions can be derived from (b) as: h + k = 2n (n is integer) for hk0, consistent with the reflection condition of $P4/nmm$ space group.} \label{fig1}
\end{figure}

\begin{figure}[ht]
\centering
\includegraphics[width=0.495\textwidth]{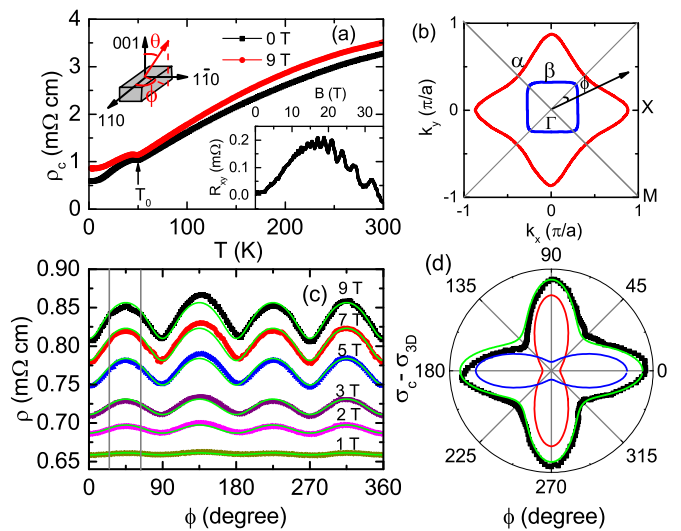}
\caption{(Color online). (a) Temperature dependence of out-of-plane resistivity of CaMnBi$_2$ in the B = 0 T and B = 9 T magnetic fields. The cartoon shows how the polar ($\theta$) and azimuthal ($\phi$) are determined. The inset show the MR of CaMnBi$_2$. (b) The Fermi surface of CaMnBi$_2$ obtained by ARPES measurements\cite{FengY}. The orientation of Fermi pocket with respect to the crystal axes are also shown. (c) The azimuthal angle ($\phi$) dependence of the out-of-plane resistivity at T = 2 K at different magnetic fields. Green lines represent fits using the three parameter equation (see text). (d) Polar plot of $\sigma_c$ -
$\sigma_{3D}$ (black square) and the three-parameter fit (green line), the red and blue lines stand for the contribution of valleys with odd and even index, respectively.} \label{fig1}
\end{figure}

The $P4/nmm$ structure of CaMnBi$_{2}$ was confirmed through neutron powder diffraction and TEM [Fig. 1(a-c)]. Neutron diffraction lattice parameters [Fig. 1(a)] are in good agreement with the reported values \cite{Brochtel}. In addition to the main phase, about 9\% by weight of elemental Bi phase was also observed due to  small amount of Bi metal flux droplets during pulverization of the single crystal specimen. TEM Electron diffraction pattern of CaMnBi$_{2}$ is also consistent with the $P4/nmm$ space group [Fig. 1(b,c)] \cite{KefengCa}.

Figure 2(a) shows the temperature dependence of the interlayer resistivity ($\rho_c$) for CaMnBi$_2$ measured at 0 T and 9 T in Quantum Design PPMS.  The overall behavior of $\rho_c$ is similar to $\rho_{ab}$.\cite{KefengCa,HeJB} Neither anomaly corresponding to the antiferromagnetic transition at $T_N$ $\cong$ 250 K nor the broad maximum at about 170 K are observed \cite{KefengCa,WangJK,GuoYF}. The broad maximum in c-axis resistivity observed before \cite{KefengCa} was probably introduced by the temperature gradient at NHMFL cryostat during the cooldown procedure. Due to antiferromagnetic order in CaMnBi$_2$, Mn-related bands are well spin-polarized and separated away from the Fermi level; consequently electronic transport properties of CaMnBi$_2$ are dominated by anisotropic Dirac cone formed by Bi $p$ band \cite{LeeG}. The origin of resistivity anomaly at $T_0$ = 50 K has been attributed to weak ferromagnetic order or spin reorientation \cite{KefengCa, HeJB}. However, this anomaly does not stem from the change of the average crystal structure since there is a smooth evolution of lattice parameters of \textit{P4/nmm} space group from 310 K to T=10 K on cooling \cite{GuoYF}. When compared to $\rho_c$ of SrMnBi$_2$,  the hump around 200 K is very weak\cite{ParkSr}, indicating that the FS is more three-dimensional (3D) than that in SrMnBi$_2$.  The magnetoresistance (MR = [$\rho_c(B)$ - $\rho_c$(0)]/$\rho_c$(0)) is about 10\% above 50 K in a 9-T field but increases strongly below 50 K to 44 \% at 2 K, similar to the in-plane MR of CaMnBi$_2$. The slope of Hall resistance R$_{xy}$(B) changes from positive to negative at $\sim$ 16 T [Fig. 2(a) inset], suggesting the presence of multiple bands in the electronic transport. According to the classical expression for the Hall coefficient when both electron- and hole-type carriers are present \cite{Allen}:

\[\begin{array}{l}
\frac{{{\rho _{xy}}}}{{{\mu _0}H}} = \mathop R\nolimits_H {\rm{ }}
 = \frac{1}{e}\frac{{(\mu _h^2{n_h} - \mu _e^2{n_e}) + {{({\mu _h}{\mu _e})}^2}{{({\mu _0}H)}^2}({n_h} - {n_e})}}{{{{({\mu _e}{n_h} + {\mu _h}{n_e})}^2} + {{({\mu _h}{\mu _e})}^2}{{({\mu _0}H)}^2}{{({n_h} - {n_e})}^2}}}
\end{array}\]

In the weak-field limit, the equation can be simplified as $R_H$ = $e^{-1}$$(\mu _h^2{n_h}-\mu _e^2{n_e})$/$(\mu _h{n_h}+\mu _e{n_e})^2$, whereas $R_H$ = 1/$(n_h-n_e)e$ in the high field limit. If $\mu_h$ $<$ $\mu_e$, the change of the slope indicates that the dominant carriers in CaMnBi$_2$ are holes at low field and electrons at high field. This is different from SrMnBi$_2$ where Hall resistance slope remains negative up to 60 T \cite{ParkSr}. The interlayer resistivity and Hall resistance suggest that the transport of CaMnBi$_2$ is dominated by 3D hole Fermi pocket at the Brillouin zone center, and that the FS is more 3D when compared to SrMnBi$_2$.

Fermi surfaces of CaMnBi$_2$ and SrMnBi$_2$ both have a hole like square-shaped parts around $\Gamma$ point. Different from the four small isolated Fermi surfaces at $\Gamma$ - M direction in SrMnBi$_2$, there is a large diamond-like FS connecting four equivalent X points in the first Brillouin zone \cite{FengY}, as shown in Fig. 2(b). The similarity of CaMnBi$_2$ and SrMnBi$_2$ Fermi surfaces indicates that valley-polarized interlayer conduction is possible in CaMnBi$_2$.

For a two fold anisotropic Fermi pocket and when the field is applied along the shorter axis, the electrons on the flat part of the FS experience nearly zero Lorentz force whereas Lorentz force makes the electrons on the side wall move along the closed orbits. This leads to minima in $\rho_c$. Therefore, magnetic field can be an effective tool to control the valley contribution to the out-of-plane resistivity \cite{ZhuZW,JoYJ}. The azimuthal angle ($\phi$) dependence of $\rho_c$ exhibits strong four fold symmetry due to the different contribution of the valleys controlled by the in-plane field orientation. For a quasi-2D FS, $\rho_c$($\phi$) can be fitted with an empirical model; we assume that the hole-like $\beta$ FS have negligible $\phi$ dependence, while the four $\alpha$ FS is ellipsoid with long axis perpendicular to $\Gamma$ - M line. Therefore, $\sigma_c$($\phi$) $\approx$ 1/$\rho_c$($\phi$) can be described by the formula \cite{JoYJ}:

\[\begin{array}{l}
{\sigma _c}(\phi ) = \sum\limits_{n = 1}^4 {{\sigma _{\alpha ,n}}(\phi )}  + {\sigma _\beta },\\
{\rm{                                                       }} = \frac{{2\mathop \sigma \nolimits_{2D} }}{{1 + r{{\cos }^2}\phi }} + \frac{{2\mathop \sigma \nolimits_{2D} }}{{1 + r{{\cos }^2}(\phi  + \pi /2)}} + \mathop \sigma \nolimits_{3D}
\end{array}\]

where $\sigma_{\rm {2D}}$ and $\sigma_{\rm {3D}}$ are the contributions of $\alpha$ and $\beta$ FS, respectively. The parameter $r$ is a measure of the anisotropy of magnetoconductivity. As shown in Fig. 1(c), all curves can be fitted with this empirical formula. The $\sigma_{\rm {2D}}$ = 0.06 (m$\Omega$ cm)$^{-1}$, $\sigma_{\rm {3D}}$ = 1.09 (m$\Omega$ cm)$^{-1}$, and $r$ = 4.28 can be obtained from the fitting of $\rho_c$($\phi$) at 9 T, and the contribution of quasi-2D FS to the $\sigma_{\rm {2D}}$ - $\sigma_{\rm {3D}}$ is illustrated in Fig. 1(d). The ratio between the quasi-2D (four $\alpha$ bands) and 3D ($\beta$ band) conductivities 4$\sigma_{\rm {2D}}$/$\sigma_{\rm {3D}}$ $\sim$ 0.22, indicating that the $\rho_c$($\phi$) is dominated by the 3D hole-like $\beta$ FS. The quasi-2D electron-like $\alpha$ bands only contribute about $\sim$ 22\% of the total out-of-plane conductivity, consistent with the small quasi-2D FSs observed by ARPES and quantum oscillations \cite{FengY,KefengCa}.

We note that the four fold symmetry is broken at high magnetic field [Fig. 1(c)] into two fold symmetry. The $\rho_c$ at $\pi$/2 and 3$\pi$/2 is larger than those at $\pi$/4 and 3$\pi$/4. Similar has been observed on SrMnBi$_2$ and Bi possibly due to the formation of nematic liquid of electrons \cite{ParkSr,ZhuZW,FradkinE}.

\begin{figure}[ht]
\centering
\includegraphics[width=0.5\textwidth]{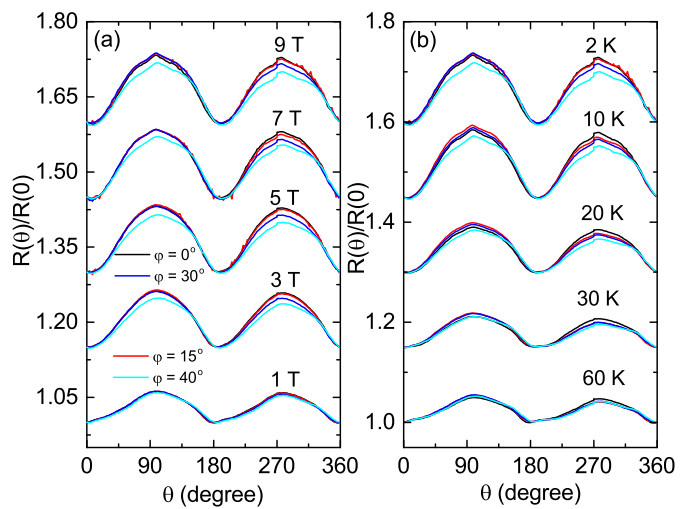}
\caption{(Color online). (a) The normalized resistivity $\rho_c$($\theta$)/$\rho_c$(0) taken at 2 K for different azimuthal angles ($\phi$) and in different fields. (b) The normalized resistivity $\rho_c$($\theta$)/$\rho_c$(0) taken at 9 T with varying ($\phi$) and temperature. (a) and (b) use same legend. Each subsequent data set are shifted upward by 0.15 for clarity.}
\label{fig2}
\end{figure}

Figure 3 shows the polar angle ($\theta$) dependence of $\rho_c$ at various azimuthal angles ($\phi$). Magnetotransport of solids is governed by the extremal cross section $S_F$ of the FS; as a result $S_F$($\theta$) = $S_0$/$\mid$cos($\theta$)$\mid$ is expected for a 2D FS. The $\rho_c$($\theta$) exhibits two fold symmetry at low temperature and high fields, and can be fitted by $\mid$cos($\theta$)$\mid$, consistent with the quasi-2D FS in CaMnBi$_2$. There are two shoulders around $\theta$ = 90$^o$, where the field is paralel to the $ab$ plane. The shoulder location (Yamaji) angles are magnetoresistance maxima where the carriers in warped cylindrical FS behave as in a 2D electronic system \cite{Ohmichi}.

The $\rho_c$($\theta$) measured in high field is presented in Fig. 4(a). It shows a peak at around 90$^{\rm o}$, which can be attributed to self-crossing orbits or closed orbits that appear on the side of the warped FS \cite{JoYJ,Hanasaki}. Moreover, the peak width is independent of the field strength, as shown in the inset of Fig. 4(a). In the case of self-crossing orbits, the angular width of the peak should be inversely proportional to the magnetic field. Thus, the peak at $\theta$ = 90$^{\rm o}$ can be ascribed to closed orbits on the side of the cylindrical FS \cite{Hanasaki}. The series of peaks between 120$^{\rm o}$ and 180$^{\rm o}$ are due to Yamaji oscillations \cite{Hanasaki}. Our results indicate the coherent interlayer conduction at low temperature and the presence of the quasi-two-dimensional(2D) FS in CaMnBi$_2$, in agreement with previous report \cite{KefengCa}.

Figure 4(b) shows the magnetic field dependence of $\rho_c$ up to 35 T. No oscillation is observed below 10 T, indicating the absence of Bi flux in MR signal since elemental bismuth shows quantum oscillations at very low magnetic fields \cite{Edelman}. The MR decreases with the increased angle. The slopes of the MRs reduce considerably at around 3 T. The resistance exhibits linear-in-field dependence at high magnetic field. No transition corresponding to the magnetic order change is observed in the resistivity at high magnetic fields. Non-saturating linear MR is has been reported in several complex materials like Ag$_{2+\delta}$Se, SrMnBi$_2$, and BaFe$_{2}$As$_{2}$ \cite{XuR,ParkSr,Huynh}. The linear MR can be observed when Dirac electrons condense at the lowest Landau Level (LLL) \cite{Abrikosov,Abrikosov2}. This is easily realized in modest magnetic fields since the distance between the LLL and first LLs of Dirac-like fermions in a magnetic field is large, in contrast to conventional parabolic-like energy dispersion \cite{Huynh,Abrikosov,ZhangY,MillerD}. Whereas the linear MR is also observed in simple metals, such as potassium\cite{Reitz}, in plane MR, quantum oscillations, thermal transport, ARPES studies and first principle calculations strongly suggest the presence of Dirac states in  CaMnBi$_2$ \cite{KefengCa,LeeG,FengY,KefengTTO}.

Shubnikov-de Haas (SdH) oscillations are observed in the c-axis electronic transport [Fig. 4(b-d)]. At first, the oscillation magnitude decreases very quickly with the angle increase, and the oscillation disappear at $\theta$ = 101$^{\rm o}$. The oscillation appear again with further angle decrease. Hence, clear oscillation can be observed when the field is either along the $c$ axis \cite{KefengCa,HeJB} or in the $ab$ plane (this study). A 3D FS would produce oscillations for all directions of magnetic field; therefore it is unlikely in CaMnBi$_{2}$ due to the absence of the oscillation at 101$^{\rm o}$. Anther possibility is that the oscillation is due to the Fermi-surface topological effect, where small closed orbits appear on the side of the warped cylindrical FS.  This model can also interpret the absence of oscillations at 101$^{\rm o}$ and the presence of the oscillations only in a narrow  range around 90$^{\rm o}$, in agreement with $\rho_c$($\theta$). As shown in the calculated Fermi surfaces of CaMnBi$_2$ in Fig. 4 in Ref. 5, this closed orbits is very likely located on the convex part of the electron Fermi pocket. The schematic of the closed orbits can be found in Fig.3(c) of the Ref. 5.

\begin{figure}[ht]
\centering
\includegraphics[width=0.45\textwidth]{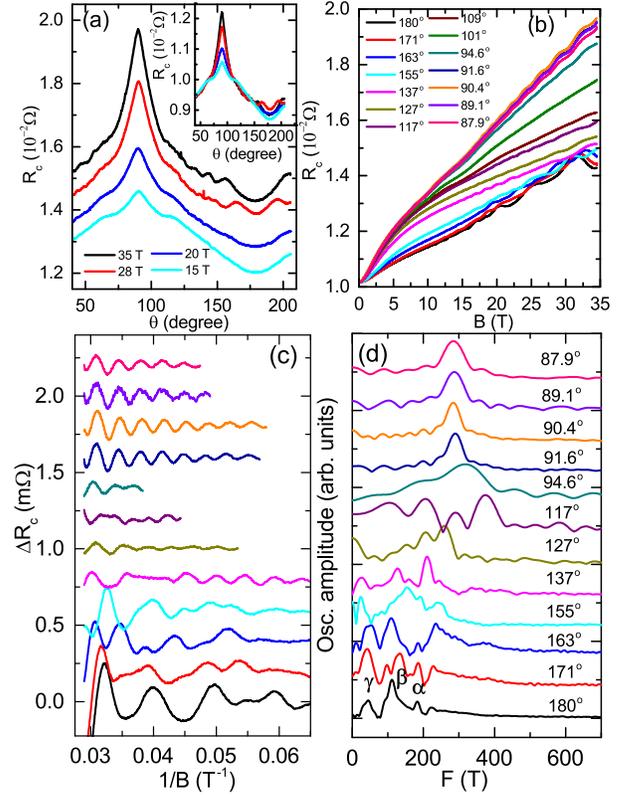}
\caption{(Color online).  (a) The angle dependence of the out-of-plane resistance R$_c$($\theta$) taken at 2 K in magnetic fields up to 35 T, inset:  R$_c$($\theta$) of CaMnBi$_2$ normalized to R$_c$(114$^{\rm o}$). (b) Magnetic field dependence of out-of-plane resistance R$_c$ up to 35 T for CaMnBi$_2$ (c) SdH oscillations component $\Delta$$R_c$ = $R_c$ - $\langle R_c \rangle$ vs 1/B for different angles measured up to 35 T at 0.7 K. (d) FFT spectra of the SdH oscillations corresponding to the oscillation component in (c). (c) and (d) use same legend. } \label{fig3}
\end{figure}

In Fig. 4(c), we show the oscillatory component of $\Delta$$R_c$ versus 1/B for different angles after subtracting a smooth background. The oscillation component shows a periodic behavior in 1/B. We perform Fast Fourier transform (FFT) on the oscillation component in Fig. 4(c), the results are presented in Fig. 4(d). There is only one frequency $F$ = 286 T when the field is parallel to the $ab$ plane.  When the field is applied along the $c$ axis, three FFT peaks are observed. The peaks are located at 45.4, 111.8, and 184.0 T, and the corresponding FS cross section is 0.43, 1.07, and 1.76 nm$^{-2}$ according to the Onsager relation $F$ = ($\Phi_0$/2$\pi^2$)$A_F$, where $\Phi_0$ is the flux quantum and $A_F$ is the cross-sectional area of the FS. All three Fermi pockets are very small, only about 0.2\%, 0.5\%, and 0.9\% of the total area of Brillouin zone. F = 184 T is very close to the dominant frequency observed before, therefore it can be assigned to $\alpha$ band with Dirac point \cite{KefengCa,HeJB}. Other frequencies could come from the $\alpha$ band which is a warped cylindrical Fermi pocket with several extremal orbits. This is consistent with the closed orbits observed on the side wall of the cylindrical Fermi pocket. The oscillations show multiband behavior when the fields tilt from the $c$ axis. In what follows we discuss the SdH when the field is applied parallel to $ab$ plane.

\begin{figure}[ht]
\centering
\includegraphics[width=0.44\textwidth]{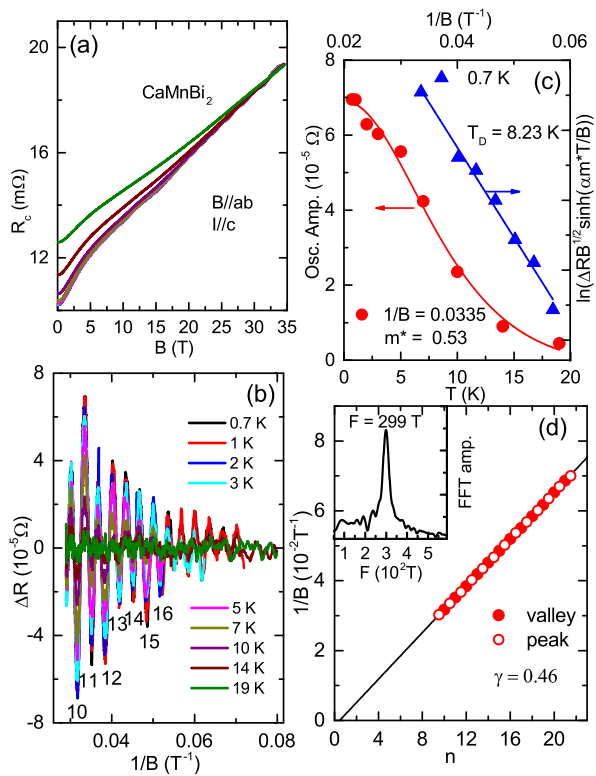}
\caption{(Color online). (a) Magnetic field dependence of resistance of CaMnBi$_2$ measured at various temperatures up to 19 K. (b) The oscillatory component as a function of 1/B. Integer indices of the Landau levels. (a) and (b) use same legend. (c) Left: temperature dependence of the oscillation amplitude at different fields. the solid lines are the fit by Lifshitz-Kosevitch formula. Right: Dingle plot at 0.7 and 1 K. (d) Landau index plots $n$ vs 1/B at 0.7 K. inset: FFT spectra at 0.7 K.} \label{fig4}
\end{figure}

Semiclassically, the SdH oscillation can be described by
\[\Delta\rho \propto {R_T}{R_D}cos[2\pi(F/B + \frac{1}{2} + \beta)]\]
where ${R}_{T} = \frac{\alpha m* T}{Bsinh(\alpha m*T/B)}$ is the thermal damping factor, and $R_D$ = exp (-$\alpha$m*$T_D$/B) is the Dingle damping factor, in which $\alpha$ = 14.69 T/K. 2$\pi$$\beta$ is the Berry's phase. Thermal damping factor can be used to determine the cyclotron effective mass from Lifshitz-Kosevitch formula. As shown in Fig. 5(c), the fitting of amplitude gives the cyclotron mass m* $\approx$ 0.53, heavier when compared to 0.35 obtained from in-plane oscillation of CaMnBi$_2$ and 0.29 in SrMnBi$_2$ \cite{KefengCa,ParkSr}. Dingle temperature $T_D$ = 8.23 K can be obtained from Dingle plot in Fig. 5(c). Therefore, scattering time of $\tau_q$ = 1.47 $\times 10^{-13}$ s can be obtained by $T_{D} = \frac{\hbar}{2\pi{k_B}\tau_q}$.  Then, the  mobility $\mu_q$ = $e\tau_q$/m$_c$ is 488 cm$^2$ V$^{-1}$ s$^{-1}$. The scattering time and mobility is even higher than in SrMnBi$_2$ and Cd$_3$As$_2$, typical Dirac materials, consistent with the presence of Dirac fermions on this Fermi pocket \cite{ParkSr,LiangT}.

The Fourier transform spectrum of the oscillation at 0.7 K reveals a periodic behavior in 1/B with a frequency $F$ = 299 T. The oscillation frequency determined by the slope of linear fit of Landau index is 299 T, in agreement with the FFT results. FS cross section normal to the field is $A_F$ = 2.86 nm$^{-2}$, and $k_F$ = 9.54 $\times 10^8$ m$^{-1}$ can be obtained. Therefore, Fermi velocity $\nu_F$ = $\hbar{k_F}/m^*$ = 2.08 $\times 10^{5}$ m/s and Fermi energy $E_F$ = 130 meV. The mean free path is estimated to be  by $l$ = $\upsilon_F$$\tau$ = 30.6 nm.

SdH oscillations in metals are related to successive emptying of LL in the magnetic field whereas the LL index $n$ is correlates with the cross section of FS $S_F$ as 2$\pi$($n + \gamma$) = $S_F$($\hbar$/$eB$). In the Landau fan diagram [Fig. 5(d)] the peaks and valleys fall on a straight line. Linear fit gives $\gamma$ $\cong$ 0.46; $\gamma$ should be zero for conventional metals but is $\pm$1/2 for Dirac fermions due to the nontrivial Berry's phase. The $\gamma$ $\sim$ 1/2 and large Fermi velocity confirm the existence of  Dirac fermions at the orbits on the side of the warped cylindrical FS. The Dirac Fermion with large Fermi velocity on the side wall of the Fermi cylinder have important effect on the $\rho_c$ which could explain why $\rho_c$ does not show clear $\phi$ dependence.

\section{CONCLUSIONS}

In conclusion, we studied the angle-dependent out-of-plane magnetotransport in CaMnBi$_{2}$.  Out-of-plane $\rho_{xx}$ and $\rho_{xy}$ indicate that the FS is more 3D when compared to SrMnBi$_2$. The interlayer conduction depends on the orientation of in-plane magnetic field. This suggests that the valley-polarized interlayer current through magnetic valley control can be realized in CaMnBi$_2$. The angular dependence of MR and the SdH show that the the closed orbits appear on the side of the warped cylindrical FS. Small FS, small cyclotron mass, large mobility, and nonzero Berry phase are consistent with the existence of Dirac fermions on the  Fermi pocket on the side wall of the warped cylindrical FS.

\section*{Acknowledgements}

Work at BNL was supported by the U.S. DOE-BES, Division of Materials Science and Engineering, under Contract No. DE-SC0012704. The experiment at ORNL Spallation Neutron Source was sponsored by the Scientific User Facilities Division, BES, U.S. DOE. Work at the National High Magnetic Field Laboratory is supported by the NSF Cooperative Agreement No. DMR-0654118, and by the state of Florida.

$^{\ast }$Present address: Department of Physics, University of Maryland, College Park, MD 20742-4111, USA.

\end{document}